\documentclass[11pt]{article}

\usepackage{amssymb}
\usepackage{amsmath}
\usepackage{graphicx}
\usepackage{hep}
\usepackage[labelfont=bf]{caption}
\usepackage{sbl_th}
\newcommand\pubblock{\rightline{\begin{tabular}{l} \pubnumber\\
 \end{tabular}}}
\newcommand\pubnumber{MPP-2008-15\\UB-ECM-PF-08/06
}
\newcommand{\msusy}{\ensuremath{M_{\mathrm{SUSY}}}}

\newcommand{\tc}{\ensuremath{tc}}
\newcommand{\bs}{\ensuremath{bs}}
\newcommand{\squark}{\ensuremath{\Psquark}}
\renewcommand{\PH}{\ensuremath{h}}
\newcommand{\higgses}{\ensuremath{\PHiggslightzero, \PHiggsheavyzero, \PHiggspszero}}
\newcommand{\bsg}{\ensuremath{\HepProcess{\Pbottom \HepTo \Pstrange \Pphoton}}}
\newcommand{\hqq}{\ensuremath{\HepProcess{\PH \HepTo \Pquark \Pquark'}}}
\newcommand{\hbs}{\ensuremath{\HepProcess{\PH \HepTo \bs}}}
\newcommand{\htc}{\ensuremath{\HepProcess{\PH \HepTo \tc}}}
\newcommand{\ppbs}{\ensuremath{\HepProcess{\Pproton\Pproton \HepTo \bs}}}
\newcommand{\pptc}{\ensuremath{\HepProcess{\Pproton\Pproton \HepTo \tc}}}
\newcommand{\ppggbs}{\ensuremath{\HepProcess{\Pproton\Pproton(\Pgluon\Pgluon) \HepTo \bs}}}

\newcommand{\pph}{\ensuremath{\HepProcess{\Pproton\Pproton \HepTo \PH}}}
\newcommand{\pphqq}{\ensuremath{\HepProcess{\Pproton\Pproton \HepTo \PH \HepTo \Pquark\Pquark'}}}
\newcommand{\pphqqD}{\ensuremath{\HepProcess{\Pproton\Pproton \HepTo\Pquark\Pquark'}}}
\newcommand{\pphtcD}{\ensuremath{\HepProcess{\Pproton\Pproton \HepTo\tc}}}
\newcommand{\pphbsD}{\ensuremath{\HepProcess{\Pproton\Pproton \HepTo\bs}}}
\newcommand{\pphbs}{\ensuremath{\HepProcess{\Pproton\Pproton \HepTo \PH \HepTo \bs}}}
\newcommand{\pphzbs}{\ensuremath{\HepProcess{\Pproton\Pproton \HepTo\PHiggslightzero  \HepTo \bs}}}
\newcommand{\pphtc}{\ensuremath{\HepProcess{\Pproton\Pproton \HepTo \PH \HepTo \tc}}}
\newcommand{\hz}{\ensuremath{\PHiggslightzero}}
\newcommand{\Pb}{\ensuremath{\Pbottom}}
\newcommand{\Pt}{\ensuremath{\Ptop}}
\newcommand{\Bbsg}{\ensuremath{B(\bsg)}}
\newcommand{\sigmapphbs}{\ensuremath{\sigma(\pphbs)}}
\newcommand{\sigmapphtc}{\ensuremath{\sigma(\pphtc)}}

\newcommand{\GIM}{Glashow:1970gm}

\newcommand{\PDG}{Yao:2006px}

\newcommand{\singletop}{LopezVal:2007rc}
\newcommand{\singletoppro}{Guasch:2006hf}

\newcommand{\Santi}{Bejar:2000ub,Bejar:2003em}
\newcommand{\Arhrib}{Arhrib:2004xu}
\newcommand{\fer}{Ferreira:2005dr}
\newcommand{\Hunter}{Hunter}
\newcommand{\separ}{\vspace{0.1cm}}

\begin{document}
\pubblock

\begin{center}
    {\Large FCNC-induced heavy-quark events at the LHC from Supersymmetry}
    \vskip 10mm
    \textbf{Santi B{\'e}jar}$^{a}$, \textbf{Jaume Guasch}$^{b,d}$,
    \textbf{David L{\'o}pez-Val}$^{c,d}$, \textbf{Joan Sol{\`a}}$^{c,d}$

    \vskip0.5cm

    $^{a}$  Max-Planck-Institut f{\"u}r Physik,\\
    F{\"o}hringer Ring 6, D--80805 M{\"u}nchen, Germany

    $^{b}$ Gravitation and Cosmology Group,  Dept. FF, Univ. de
    Barcelona\\  Av. Diagonal 647, E-08028
    Barcelona, Catalonia, Spain \\

    $^{c}$ High Energy Physics Group, Dept. ECM, Univ. de Barcelona\\
    Av. Diagonal 647, E-08028 Barcelona, Catalonia, Spain

    $^{d}$ Institut de Ci{\`e}ncies del Cosmos, UB, Barcelona.

    E-mails: sbejar@mppmu.mpg.de, jaume.guasch@ub.edu, dlopez@ecm.ub.es,
    sola@ifae.es.

    \vskip2mm

\end{center}
\vskip 15mm

\begin{abstract}
We analyze the production and subsequent decay of the neutral Higgs
bosons $\PH\equiv \higgses$ of the MSSM into electrically neutral
quark pairs of different flavors ($qq'\equiv \tc,\ \bs$, depending
on $h$) at the LHC, i.e. $\sigma(\pphqq)$, and compare  with the
direct FCNC production mechanisms $\sigma(\pphqqD)$. The
cross-sections are computed in the unconstrained MSSM with minimal
flavor-mixing sources and taking into account the stringent bounds
from $\bsg$. We extend the results previously found for these FCNC
processes, which are singularly uncommon in the SM. Specifically, we
report here on the SUSY-EW part of $\sigma(\pphqq)$ and the SUSY-QCD
and SUSY-EW contributions to $\sigma(\pphbsD)$. In this way, the
complete map of MSSM predictions for the $qq'$-pairs produced at the
LHC becomes available. The upshot is that the most favorable
channels are: 1) the Higgs boson FCNC decays into $\bs$, and 2) the
direct production of $\tc$ pairs, both of them at the $\sim
1\picobarn$ level and mediated by SUSY-QCD effects. If, however, the
SUSY-QCD part is suppressed, we find a small SUSY-EW yield for
$\sigma(\pphtc)_{\rm max}\sim 10^{-4}\picobarn$ but, at the same
time, $\sigma(\pphbs)_{\rm max}\sim 0.1-1\,\picobarn$, which implies
a significant number ($\sim 10^4-10^5$) of $\bs$ pairs per
$100\,\invfb$ of integrated luminosity.
\end{abstract}




\section{Introduction}
\label{Introduction} The upcoming generation of $\TeV$-class
colliders, headed by the imminent startup of the Large Hadron
Collider (LHC) at CERN, will offer us the opportunity to dig deeper
than ever into the nature of fundamental interactions. Despite its
successful career, the Standard Model (SM) of elementary particles
still offers a number of intriguing puzzles to be resolved, such as
the ultimate origin of the electroweak (EW) symmetry breaking and
the mass generation mechanism. Among the wide set of theoretical
proposals that have been conjectured so far, the possibility that
the fundamental laws of Nature exhibit a symmetry between fermionic
and bosonic degrees of freedom -- that is, supersymmetry (SUSY) and
most particularly the Minimal Supersymmetric Standard Model
(MSSM)~\cite{MSSM} -- has been postulated as one of the firmest
candidates to account for possible scenarios of physics beyond the
SM, i.e. of New Physics (NP).

The quest for experimental signatures of an underlying SUSY dynamics
is undoubtedly one of the major endeavors for the LHC. Besides the
direct production and tagging of SUSY particles (which could be
cumbersome in practice), it is of great interest to consider the
impact of SUSY radiative corrections on conventional processes in
the SM. In such cases the enhancement capabilities associated to the
non-standard couplings may provide sources of genuine NP signatures.
This possibility was thoroughly analyzed in the past for the $W$ and
$Z$ boson physics (see e.g. \cite{OldSUSYJS,Chankowski:1993eu}).

More recently, a great deal of attention has been devoted to
processes triggered by Flavor-Changing Neutral-Current (FCNC)
sources beyond the SM, in particular from SUSY interactions. From
the seminal work of Glashow, Iliopoulos and Maiani (GIM) \cite{\GIM}
it has been known that the FCNC interactions are absent at the tree
level within the SM and,  most significantly, they are largely
suppressed at the $1$-loop order. This, so-called, GIM mechanism is
an in-built feature of the mathematical core of the SM, thanks to
the unitarity of the CKM matrix. Low energy $\PB$-meson physics, for
example, provides a number of well-measured FCNC processes, such as
the $\PB$-meson radiative decay $\bsg$. Its branching ratio reads
 $\Bbsg = (2.1 - 4.5)\times 10^{-4}$ at the $3\sigma$ level\,
\cite{Yao:2006px}. This important and well studied process can be
used to constrain models of NP that predict non-standard
flavor-changing interactions. In contrast, the FCNC effects
involving the top quark as an external particle turn out to be
dramatically suppressed by the GIM mechanism. The predicted
branching ratios are at the level of $B(\HepProcess{\Ptop\HepTo
\Pcharm \Pgluon})\sim 10^{-11}$ and $B(\HepProcess{\Ptop\HepTo
\Pcharm H})\sim 10^{-14}$\,\cite{Diaz-Cruz:1989ub}, hence far below
the limits of observability.

In stark contrast with this meager SM panorama, the MSSM opens new
vistas for a fruitful FCNC physics of the top quark. A most
remarkable example is the case of the top quark decay into the
lightest supersymmetric neutral Higgs boson, $\PHiggslightzero$,
whose branching ratio could be of order $B(\HepProcess{\Ptop \HepTo
\Pcharm \PHiggslightzero })\sim
10^{-4}-10^{-3}$\,\cite{Guasch:1999jp} -- previously underestimated
in \cite{Yang:1993rb}. A branching ratio of this order represents
not only an enhancement of $10$ orders of magnitude above the SM
prediction, but it opens the realistic possibility for measurement.
The origin of these possible effects lies in the richer diversity of
sources of flavor mixing in the MSSM, in particular among fermion
and sfermions of the same charge and different generation. They
ultimately stem from the so-called misalignment of the squark mass
matrices with respect to the quark ones and can be described by
means of the parameters $\delta_{ij}^{AB}$ (being $A,B=L,R$ the
chirality indices and $i,j=1,2,3$ the flavor ones). A generic entry
of the soft SUSY-breaking squark mass matrix reads $(M^2)^{AB}_{ij}
= \delta_{ij}^{AB}\,\tilde{m}_i\,\tilde{m}_j$ (for $i\neq j$). As a
consequence, new types of FCNC couplings arise and they need not be
subdued by GIM suppression.
 Obviously, such scenarios bring us some
hope to effectively unearth hints of SUSY physics out of the study
of both low energy and high energy FCNC
processes\,\cite{Misiak:1997ei}.

Interestingly enough, not only SUSY can help here; other alternative
proposals of extended physics beyond the SM, among them the general
Two-Higgs-Doublet Models (2HDM) \cite{\Hunter}, topcolor models and
strong flavor-changing effects predict in some cases an enhanced,
and sometimes distinctive, FCNC phenomenology
\cite{\Santi,\Arhrib,\fer}. Attentive studies on this field can thus
be of great help in seeking for new signatures and eventually
disentangling the sort of NP hiding right there.\footnote{For a
review of top quark and Higgs boson FCNC physics in the MSSM and in
general 2HDM models, see e.g.~\cite{Bejar:2001sj} and references
therein.}

In this letter, we wish to further elaborate on the rich
phenomenology of flavor-changing processes in the MSSM, specifically
on those that could lead to an unsuspected overproduction rate of
electrically-neutral heavy-quark pairs $qq'$ of different flavors at
the LHC. Most conspicuously, we have the process of single top quark
production induced by FCNC, i.e. any of the two-body final states
$\pptc\equiv \Ptop\APcharm + \APtop\Pcharm$. This process is
extremely suppressed in the SM, while it can be highly enhanced in
the MSSM\,
\cite{Liu:2004bb,Guasch:2006hf,Eilam:2006rb,Cao:2006xb,LopezVal:2007rc}.
A similar situation applies to producing FCNC single b-events, i.e.
$\bs\equiv\Pbottom\APstrange + \Pstrange\APbottom$, although here
the SM suppression is not so drastic as in the $\tc$ case.

There are different supersymmetric mechanisms leading to enhanced
$\tc$ and $\bs$ final states. As a first possibility, we have the
SUSY flavor-changing charged (and neutral) currents contributing to
the direct production processes $\pptc$ and $\ppbs$. An alternative
mechanism to produce these final states is through the production
and subsequent FCNC decay of a neutral Higgs boson, i.e. $\pphtc$
and $\pphbs$. This mechanism was explored within the context of the
general 2HDM \cite{Bejar:2003em}, in which $\PH =
\higgses$\,\cite{\Hunter}, and it was later revisited within the
unconstrained MSSM through the strong supersymmetric (SUSY-QCD)
corrections to the Higgs boson FCNC branching ratios into the
heavy-quark final states \cite{Curiel:2002pf,Bejar:2004rz}. However,
a first computation of the corresponding LHC production rates
appeared only in\,\cite{Bejar:2005kv}.

Supersymmetric electroweak (SUSY-EW) contributions to neutral Higgs
boson decays could be important too. These effects can be induced,
in principle, by both charged and neutral currents. The latter are
triggered by the neutralinos and some authors have dealt with these
effects ~\cite{Demir:2003bv,Curiel:2003uk,Hahn:2005qi}. Here,
however, following \cite{Guasch:1999jp} (see also
\cite{Arhrib:2006vy}), we compute the SUSY-EW contributions in the
super-CKM basis, i.e. under the assumption of minimal flavor
violation ($\delta_{ij}^{AB}=0,\ i\neq j$). We argue and verify that
the neutralino effects are subleading. Therefore, we concentrate on
the charged current sector of the electroweak MSSM
Lagrangian\,\cite{DLV:I07}. In this setup, we perform a systematic
analysis along the lines of \cite{Bejar:2005kv} by addressing the
SUSY-EW effects on $\sigma(\pphqq)$ induced by charginos, squarks
and charged Higgs bosons in the MSSM. We also compute the pay-off in
the number of $\bs$ events from the direct production mechanism
$\sigma(\pphbsD)$ within the MSSM, including both SUSY-QCD and
SUSY-EW effects. Notice that the interest of knowing the SUSY-EW
effects in these studies is that they could be the only sizeable
supersymmetric source of FCNC $qq'$-pairs at the LHC, if gluinos
turn out to be very heavy and/or the inter-generational mixing
parameters would be too small or simply zero\,\footnote{When
considering the explicit FCNC effects on our processes, we just
focus on $\delta_{23}\equiv\delta_{23}^{LL}$. We assume that the
FCNC mixing may hold in the chiral $LL$ sector of the squark
$6\times 6$ mass matrices only, which is well motivated
theoretically. The other chirality sectors give similar
contributions\,\cite{Guasch:1999jp}. See also
\cite{Liu:2004bb,Eilam:2006rb}.}.

\section{Higgs boson decays into neutral heavy quark-pairs}
\label{sec:higgs-mediated}

In this section, we discuss the general expectations on the
production cross-section of electrically neutral pairs of heavy
quarks of different flavors at the LHC, whose origin stems from the
FCNC decays of a neutral MSSM Higgs boson \cite{Hunter}. The main
contribution to the overall process may be sequentially split into:
i) production of a real neutral Higgs boson in a proton-proton
collision, ii) followed by its decay through loop diagrams
contributing to the FCNC final state. Assuming that the Higgs bosons
are produced on-shell, the total production rate can be factorized
as follows:
\begin{eqnarray}
        \sigma(\pphqq)
        &\equiv&
        \sigma(\pph X)B(\hqq)\nonumber\\
        &\equiv&\sigma(\pph X)
            \frac{\Gamma(\hqq)}{\Gamma(\HepProcess{\PH\HepTo Y)}}\ \ \
        (qq'\equiv bs \mbox{ or } tc )\,.
    \label{eq:hqq-def}
\end{eqnarray}
Here $\Gamma(\hqq)$ is the total FCNC two-body partial decay width
of the corresponding MSSM Higgs boson $h=\higgses$ into the
(kinematically possible) neutral states $\bs\equiv\Pbottom\APstrange
+ \Pstrange\APbottom$ or $\tc\equiv \Ptop\APcharm + \APtop\Pcharm$;
and $\Gamma(\HepProcess{\PH\HepTo Y})$ stands for the --
consistently computed -- total decay width in each case.
\begin{figure}[tp]
\begin{center}
        \includegraphics*[height=0.30\textwidth]{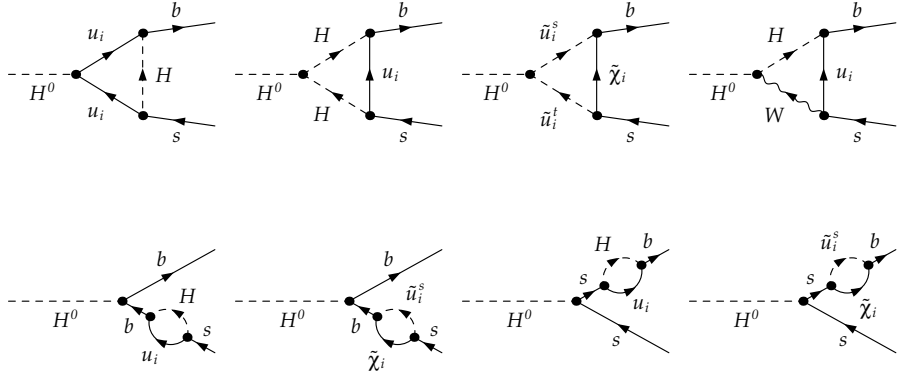}
    \caption{Sample of SUSY-EW contributions to the decay of the Higgs boson
$\PHiggsheavyzero$ into $\bs$ from the charged current. The
electroweak effects that could originate from the neutralinos are
not shown. A similar collection of diagrams describes the
corresponding decay into $\tc$. } \label{fig:diag}
\end{center}
\end{figure}

Before coming to grips with the full numerical analysis, it is
always a good exercise to perform an analytical estimate based on
physical considerations. In doing this, we expect to gain insight
into the dominant FCNC sources of enhancement within the MSSM and
roughly predict (within order of magnitude, hopefully) the relation
between the cross-sections $\sigma(\pphtc)$ and $\sigma(\pphbs)$. In
order to address such estimate, we first need to specify the
possible supersymmetric electroweak contributions to the FCNC Higgs
boson decays into the heavy-quark final states under consideration.
For the sake of definiteness, let us concentrate on
$\PHiggsheavyzero$ (i.e. the heavy CP-even Higgs boson of the MSSM).
A sample of Feynman diagrams from the SUSY-EW sector is displayed in
Fig.~\ref{fig:diag}. The overall set is, of course, UV finite (as we
have checked). However, we note that the subsets of charged gauge
and Higgs boson mediated diagrams, on the one hand, and the chargino
mediated, on the other, are separately UV-finite. Moreover, each of
these subsets involves two different topologies, vertex corrections
(V) and wave-function (WF) renormalization, whose respective
contributions are also separately UV-finite after summation over
generations.

It turns out that the chargino-mediated pieces drive the bulk of the
contribution, basically because they are not suppressed by the GIM
mechanism. Let us first consider the WF
corrections. Using no other tools than dimensional analysis, power
counting, CKM matrix elements and dynamical features, we can
estimate the (one-loop) decay width of
$\HepProcess{\PHiggsheavyzero \HepTo \bs}$ as follows:
\begin{eqnarray}
\Gamma_{\bs}^{\rm WF} &\sim& m_{H}\,G_F^3\,
\left(\frac{m_t^2\,m_b}{M^2_{SUSY}}\right)^2\,
\left(\frac{V_{ts}\,V_{tb}^*}{16\pi^2}\right)^2\,\left(A_t
-\frac{\mu}{\tan\beta}\right)^2
\left(\frac{\mu\,\cos\alpha}{\cos^2\beta\,\sin\beta}\right)^2
\label{eq:gamma_cha_bs1}\, ,
\end{eqnarray}
and similarly for the $\HepProcess{\PHiggsheavyzero \HepTo \tc}$
channel:
\begin{eqnarray}
\Gamma_{\tc}^{\rm WF} &\sim& m_{H}\,G_F^3\,
\left(\frac{m_t\,m_b^2}{M^2_{SUSY}}\right)^2\,
\left(\frac{V_{tb}\,V_{bc}^*}{16\pi^2}\right)^2\,\left(A_b
-\mu\tan\beta\right)^2
\left(\frac{\mu\,\sin\alpha}{\cos\beta\,\sin^2\beta}\right)^2
\label{eq:gamma_cha_tc1}.
\end{eqnarray}
Let us briefly explain the origin of the different terms appearing
in the expressions above. In both cases we include the standard
numerical factor arising from the loop integration. $G_F$ stands for
the Fermi constant, and $V_{tb}\sim 1, V_{bc}\sim 0.02, V_{ts} \sim
0.02$ are the CKM matrix elements; $M_{SUSY}$ defines a common scale
for the soft SUSY-breaking masses (of squarks and gauginos), the
MSSM trilinear couplings are labelled by $A_b, A_t$; and $\mu$
indicates the higgsino mass parameter. {The latter accounts for the
helicity flip that appears along the chargino line\footnote{For an
explanation and a computation of the helicity flips
  in the quark propagator, see e.g. Refs.\cite{Coarasa:1996qa},
  specifically Fig. 7 and eq.~(76) of the first work
  in~\cite{Coarasa:1996qa}.}.
The angle $\alpha$ is the mixing angle between the neutral CP-even
Higgs boson states. Finally, $\tan\beta \equiv v_2/v_1$ is the ratio
of vacuum expectation values of the two MSSM Higgs
doublets\,\cite{Hunter}. A key element in the expressions above is
the factor associated to the chiral transition of top squarks,
$m_t\,(A_t-\mu\tan\beta)$, or bottom squarks,
$m_b\,(A_b-\mu\tan\beta)$, depending on the case. In the estimate,
we also include the Higgs-quark-quark and (the leading part of) the
chargino-quark-squark couplings~\cite{Hunter}, from which we get the
remaining dependences in $\alpha, \beta$. Last but not least, we
also keep track of the relevant mass scales arising from the 2-point
loop functions and the phase space integration, which we settle in
terms of $M_{SUSY}$ and the mass of the decaying Higgs particle,
$m_{H}$. Note that equation~(\ref{eq:gamma_cha_bs1}), for example,
patently reveals different sources of possible large enhancements,
mainly associated with large values of the trilinear coupling $A_t$,
the higgsino mass parameter $\mu$ and/or the trigonometric factor
$1/\cos^4\beta$ (which behaves as $\sim \tan^4\beta$ in the large
$\tan\beta$ regime).

Similar arguments can be applied to the vertex corrections, and we
arrive at
\begin{eqnarray}
\Gamma_{\bs}^{\rm V} &\sim& m_H\, G_F^3\,
\left(\frac{V_{tb}^*\,V_{ts}}{16\pi^2}\right)^2
\left(\frac{m_t^2\,m_b}{M_{SUSY}^2}\right)^2
\left[\frac{\mu\,(A_t\sin\alpha
-\mu\cos\alpha)}{\sin^2\beta\,\cos\beta}\right]^2,
\label{eq:gamma_cha_bs2} \\
\Gamma_{\tc}^{\rm V} &\sim& m_H\, G_F^3\,
\left(\frac{V_{tb}\,V_{bc}^*}{16\pi^2}\right)^2
\left(\frac{m_t\,m_b^2}{M_{SUSY}^2}\right)^2
\left[\frac{\mu\,(A_b\cos\alpha -
\mu\sin\alpha)}{\sin\beta\,\cos^2\beta}\right]^2
\label{eq:gamma_cha_tc2}.
\end{eqnarray}
Some differences between the formulae above corresponding to $\tc$
and $\bs$ final states are worth noticing. For instance, the quark
(or squark) mass insertions involve distinct mass factors: $m_b$ (in
the $\tc$ case) and $m_t$ (in the $\bs$ one). The trigonometric
couplings are also different and for this reason one of the channels
may be much more suppressed than the other at different regimes. In
particular, we see that in both cases the decay rate increases with
$\tb$, but while the leading effect of the $\bs$ channel lies in the
WF corrections the dominant one in the $\tc$ channel resides in the
V contribution.  As for the neutralino yield (whose diagrammatic
effects are not shown in Fig.\,\ref{fig:diag}), similar analytical
estimates can be produced. We limit ourselves to single out some
differences. To start with, we emphasize that these contributions
are proportional to the inter-generational mixing parameters and
are, therefore, vanishing for $\delta_{ij}^{AB}=0\,(i\neq j)$. The
corresponding effects in the WF sector are such that e.g. the
factors $A_b -\mu\tan\beta$ and $A_t -\mu/\tan\beta$ become
interchanged in the counterparts of Eqs.
(\ref{eq:gamma_cha_bs1})-(\ref{eq:gamma_cha_tc1}). Similarly with
the factors $A_t\sin\alpha -\mu\cos\alpha$ and $A_b\cos\alpha -
\mu\sin\alpha$ in the corresponding vertex contributions, i.e., the
analogous of Eqs. (\ref{eq:gamma_cha_bs2})-(\ref{eq:gamma_cha_tc2}).

In sections~\ref{sec:analysis-bs}-\ref{sec:analysis-tc}, we will
come back to the previous analytical estimates and shall compare
them with the exact numerical results. This will be useful to track
the dynamical origin of the leading MSSM contributions and, in
particular, to argue that the neutralino effects are negligible.

Similar considerations can be done for the other Higgs bosons.
However, we recall that $\PHiggslightzero$ (the lightest CP-even
Higgs boson) cannot decay into $tc$ because
$m_{\PHiggslightzero}<m_t$ in the MSSM. Moreover, for this Higgs
boson, the implications of the so-called ``small $\alpha_{eff}$''
scenario (triggered by large radiative corrections in the parameters
of the MSSM Higgs sector) must be taken into account
\cite{Carena:1999bh}.

\section{Framework for the numerical analysis}
\label{sec:hbsSUSY_prod:numerical-analysis}

In order to compute the SUSY-EW one-loop diagrams contributing to
the FCNC cross-sections $\sigma(\pphqq)$ for the three MSSM neutral
Higgs bosons ($\PH= \higgses$), we shall closely follow the notation
and methods of
Refs.\,\cite{Bejar:2004rz,Bejar:2003em,Bejar:2000ub,Guasch:1999jp,Bejar:2005kv}.
We address the reader to these references for the technical details.
In particular, a thorough exposition of the relevant interaction
Lagrangians and similar set of Feynman diagrams for the FCNC
interactions, is provided in \cite{Guasch:1999jp}. See also
Ref.~\cite{Hunter} for basic definitions in the MSSM framework and
\cite{Coarasa:1996qa} for detailed computational techniques and
further illustration of the supersymmetric enhancement effects in
other relevant Higgs boson processes. Along our computation we have
made use of \texttt{HIGLU}, \texttt{PPHTT}~\cite{Spira},
\texttt{LoopTools}, \texttt{FeynArts} and
\texttt{FormCalc}~\cite{Hahn:1998yk}.

For the numerical evaluation, we shall adhere (wherever possible) to
the general procedure put forward in Ref.~\cite{Bejar:2005kv}, where
significant parts of the numerical contributions to the
cross-sections have already been reported. Next we specify the
framework under which the computation of the observables
(\ref{eq:hqq-def}) has been carried out in the present work:
\begin{itemize}
\item We compute the one-loop SUSY-EW contributions to the FCNC partial
    decay widths $\Gamma(\hqq)$ and compare with the corresponding
    SUSY-QCD effects\,\cite{Bejar:2005kv};
\item The SUSY-EW part is defined to be
the set of contributions from charginos, neutralinos and Higgs
bosons. In practice, however, we will provide the detailed results
from the charged current effects, and argue (and numerically verify)
that the neutralino contributions are comparatively negligible.
\end{itemize}
Our major goal is to assess what are the theoretical expectations on
the observables (\ref{eq:hqq-def}) within the MSSM and, most
particularly for the present work, to ascertain whether in the
absence of strong supersymmetric sources of FCNC the electroweak
SUSY sector can still provide significant rates. To this aim we have
performed a systematic scan all over the MSSM parameter space and
have determined the maximum values for the production
rates~(\ref{eq:hqq-def}) under study. Furthermore, in order to keep
the CPU-time under a feasible range, such a scan has been carried
out by means of a Monte Carlo sampling method\,\cite{Brein:2004kh}
based on the well-known Vegas integration
program\cite{Lepage:1977sw}. This numerical procedure was amply
tested in the similar computation presented in
Ref.~\cite{Bejar:2005kv}.

On this basis, we have performed a maximization of the FCNC
cross-section within the following restrictions:
\begin{equation}
    \begin{array}{c|c|c}
        \hline
        \tan\beta & \multicolumn{2}{c}{50} \\
         A_t&\multicolumn{2}{c}{|A_t|\leq3\msusy}\\
         A_b&\multicolumn{2}{c}{|A_b|\leq3\msusy}\\
         \mu&\multicolumn{2}{c}{(0 \cdots 1000)\,\GeV}\\
         m_{\tilde{q_i}}&\multicolumn{2}{c}{\msdl=\msdr=\msur=\mg\equiv\msusy}\\
         M_2 & \multicolumn{2}{c}{\msusy}\\
         \msusy&\multicolumn{2}{c}{(150 \cdots 1000)\, \GeV}\\
         \mA& \multicolumn{2}{c}{(100\cdots 1000)\,\GeV}\\
         M_{\tilde{q_i}}& \multicolumn{2}{c}{2\,M_{\tilde{q_i}}>\mH+ 50\,\GeV}\\
        &\multicolumn{2}{c}{M_{\tilde{q}_i}+M_{\tilde{q}_j}>
          \mA+50\,\GeV \, \, (i\neq j)}\\
        \Bbsg & \multicolumn{2}{c}{(2.1-4.5)\times 10^{-4}\quad(3\sigma)}\\
        \hline
    \end{array}
    \label{eq:scan-parameters}
\end{equation}
Here $m_{\tilde{q_i}}=m_{\tilde{q}_{L,R}}$ are the squark soft
SUSY-breaking mass parameters in each chiral sector, which are
common for the three generations; $M_{\tilde{q_i}}$ are the physical
masses of the squarks, $M_2$ is the $SU(2)_L$ gaugino mass, and
$\Mh$ stands for the mass of the corresponding Higgs boson
$\PH=\higgses$. Due to the structure of the couplings in the MSSM,
and taking into account that the range $\tb\lesssim 2.5$ is not
favored in the MSSM, the parameter $\tb$ is fixed at a high value
for both channels as indicated. Concerning the characteristic SUSY
mass parameter $\msusy$, it sets the scale for the masses of the
squarks and gauginos. Notice also that our analysis incorporates a
specific choice of sign for $\mu$ ($>0$) partially motivated by the
data on the muon anomalous magnetic moment\,\footnote{The sign
$\mu>0$ is not essential for our numerical results. Moreover, the
observable $g_{\mu}-2$ also depends on the value of some slepton
masses which do not play any role in our calculation. For the SUSY
effects on $g_{\mu}-2$, see e.g. the excellent review
\cite{Stockinger:2006zn} and references therein.}. We adopt
$\Bbsg=(2.1-4.5)\times 10^{-4}$ as the experimentally allowed range
at the $3\sigma$ level~\cite{\PDG} and include in our codes the MSSM
computation of the branching ratio at leading order from
Ref.~\cite{Bobeth:1999ww}. In addition, we make sure that the sign
of the MSSM amplitude for $\bsg$ and the purely-SM one do
coincide\,\cite{Gambino:2004mv}. By enforcing these experimental and
theoretical constraints in our calculation, we automatically eschew
regions of the MSSM parameter space which would artificial enhance
the predicted $qq'$ rates at the LHC.

\separ Concerning the numerical computation of the direct $\bs$
production $\ppbs$ (mainly from gluon-gluon fusion $\ppggbs$), some
technical stumbling blocks have to be overcome. {On the one hand, it
is well-known that the presence of light quarks in the final state
of such type of processes entails large logarithmic factors, which
depend on the soft scales of the problem -- precisely the masses of
these light quarks. These large logarithms are the remnants of the
truly collinear divergences that would arise if quarks were
massless, and they turn out to be related to the non-perturbative
regime of QCD. Upon resummation, these terms can be factored out
from the computation of the partonic cross section, and finally
reabsorbed into the definition of the parton fragmentation
functions. Nevertheless, we do not need to address such a detailed
analysis here. Instead, we can include an angular cut ($\sin^2
\theta \ge$ 0.05) to circumvent such delicate collinear scenario.
Such a simpler strategy should be enough to attest the fact that the
direct production of $\bs$ is, by far, a subleading mechanism -- See
Section~\ref{sec:analysis-bs} for details.}

A second (and even more subtle) difficulty
is caused because the $\bs$ production threshold is very small. In
the limit of very low external momenta (which is, by the way, the
situation when we probe partonic $\sqrt{S}$ energies near the $\bs$
threshold), we encounter one-loop expressions of the sort
\begin{equation}
\mathcal{M} \sim \int\,d^4\,q\,\frac{f(q)}{((q+\epsilon)^2-m_1^2)\
(q^2 -m^2_1)}\, \dots \label{eq:e1}
\end{equation}
where $\epsilon\ll m_1$. This kind of expression, when evaluated
numerically, generates pseudo-singularities of the kind
$\mathcal{O}(1/\epsilon^2)$ in some of the intermediate steps, a
fact that is reflected in the associated integration uncertainties.
Numerical instabilities have a critical impact on the overall
computation, which involves a large number of diagrams (some of them
including various ``box diagrams'', i.e. $4$-point functions). We
refrain from displaying here the full list of diagrams, which is
very similar to those shown in Figs.\,1-5 of
Ref.\cite{LopezVal:2007rc}, after appropriately replacing the
virtual quark and squark contributions in accordance with the new
external $b$ and $s$ states. Particularly subtle are the
instabilities involved in solving the Passarino-Veltman reduction
formulae necessary to obtain the $4$-point amplitudes (a cumbersome
diagonalization procedure which is extremely sensitive to numerical
niceties).

A thorough analysis of such instabilities has been performed, from
which we conclude that we can handle them by: i) changing the
subroutine that undertakes the Passarino-Veltman reduction for the
$4$-point amplitudes within the LoopTools framework -- we use,
instead, an independent version presented in \cite{Denner:1991kt},
which is also implemented in the LoopTools code but only used to
cross-check the results in the standard setup. This way the problem
is smoothed, albeit not fully solved since the new subroutine
renders lower uncertainties in such delicate regimes; and ii)
introducing a cut over the partonic integration domain in order to
elude the neighboring region of the $\bs$ threshold where the
numerical instabilities arise.

\section{Numerical analysis of the $\bs$ production rate}
\label{sec:analysis-bs}

Let us first consider the Higgs boson production/decay mechanism
$\pphbs$. The main result of our Monte Carlo scan is shown in
Fig.~\ref{fig:hbs_prod_ma}a, which displays the maximum values of
the production cross-section $\sigmapphbs$, Eq.\,(\ref{eq:hqq-def}),
for the three MSSM Higgs bosons $h=\higgses$ at the LHC as a
function of $\mA$. On the left-vertical axis of this figure, we
indicate the value of the cross-section (in pb) and at the same time
we track the number of FCNC events (per $100\;\invfb$ of integrated
luminosity $\int{\cal L}\,dt$) on the right vertical axis. One can
immediately see that, in the large $\tan\beta$ regime: i) The
maximum number of events reaches a sizeable level, which lies
between $10^4$ and $10^5$ per $\int{\cal L}\,dt=100\;\invfb$ in the
interesting mass region $100\,\GeV\lesssim \mA\lesssim 200\,\GeV$.
In the particular case of the $\PHiggslightzero$ channel, the rate
of $\sim 10^4$ events extends even farther for larger $\mA$ (up to
around $250\,\GeV$); ii) Actually, for this channel, there is an
approximate plateau extending across the mass range
$300\,\GeV\lesssim\mA\lesssim 600\,\GeV$, where the event rate $\sim
10^3$ is sustained; iii) The other two channels $\PHiggsheavyzero,
A^0$ can reach a similar (though smaller) maximum number of events,
but only in the strict range $100\,\GeV\lesssim \mA\lesssim
200\,\GeV$ beyond which the rate decreases steadily and unstoppably.
We can compare this behavior with the SUSY-QCD
case\,\cite{Bejar:2005kv} -- see Fig.~\ref{fig:hbs_prod_ma}b --
which looks similar but significantly shifted upwards.
\begin{figure}[tp]
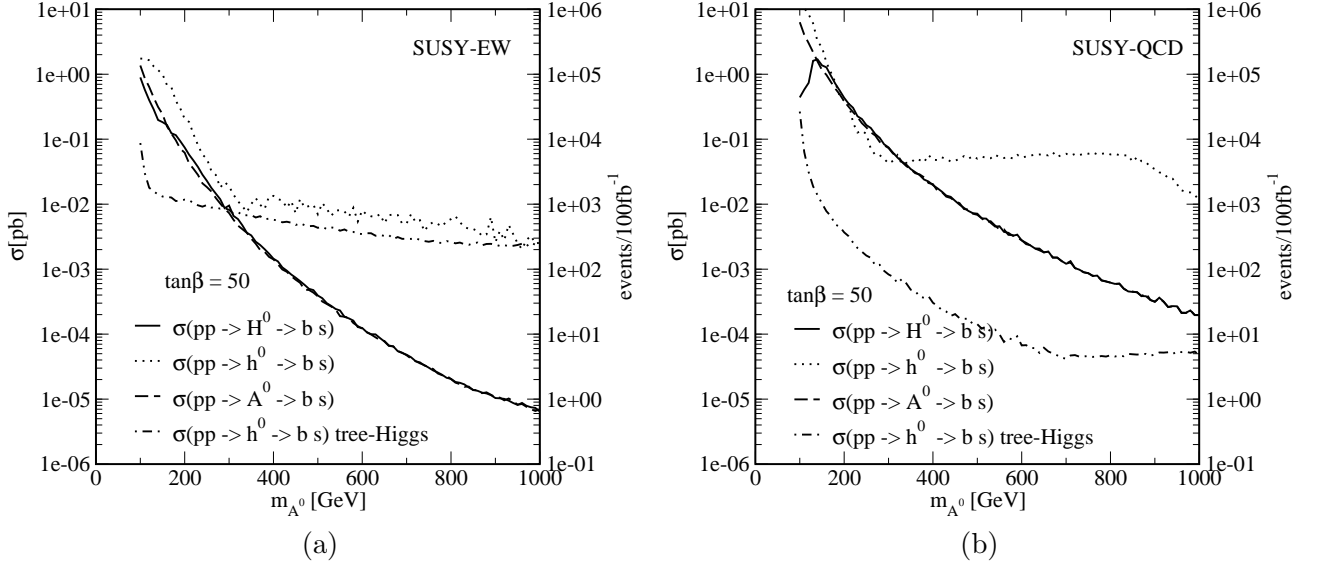

    \begin{tabular}{cc}
        \includegraphics*[height=0.42\textwidth]{hbs_prod_pphqq_ma} &
        \includegraphics*[height=0.42\textwidth]{hbsSQCD_prod_pphqq_ma} \\
        (a) & (b)
    \end{tabular}
    \caption{Maximum contributions to $\sigmapphbs$ in
      Eq.\,(\ref{eq:hqq-def}) as a function of
      $\mA$ (at fixed $\tb$) taking into account: \textbf{a}) SUSY-EW interactions
      with minimal flavor-mixing; \textbf{b}) SUSY-QCD interactions (see
      Ref. \cite{Bejar:2005kv}). The
      left-vertical axis provides the cross-section (in pb) and the
      right-vertical axis tracks the number of events per $100\;\invfb$
      of integrated luminosity.}
    \label{fig:hbs_prod_ma}
\end{figure}
\begin{figure}[tp]
    \begin{center}
\begin{tabular}{cc}
        \includegraphics*[height=0.42\textwidth]{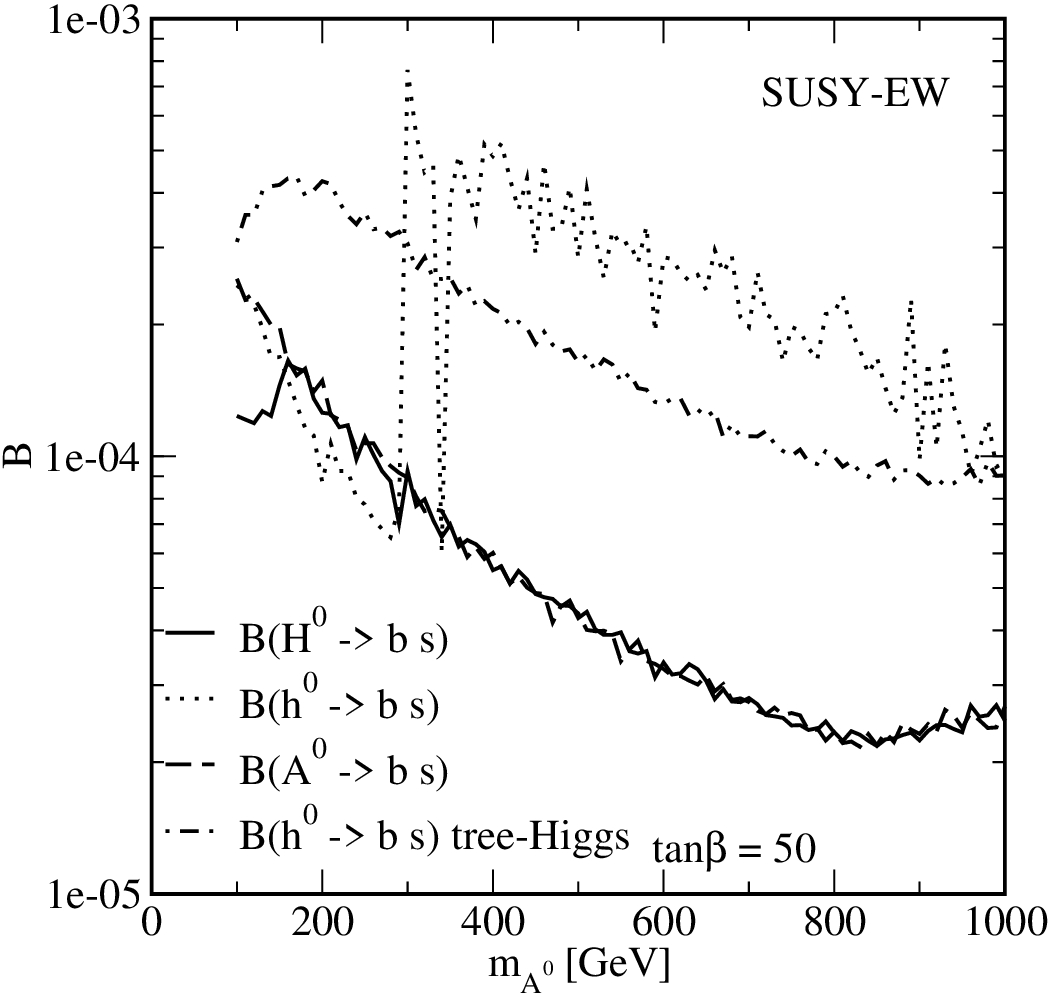} &
        \includegraphics*[height=0.42\textwidth]{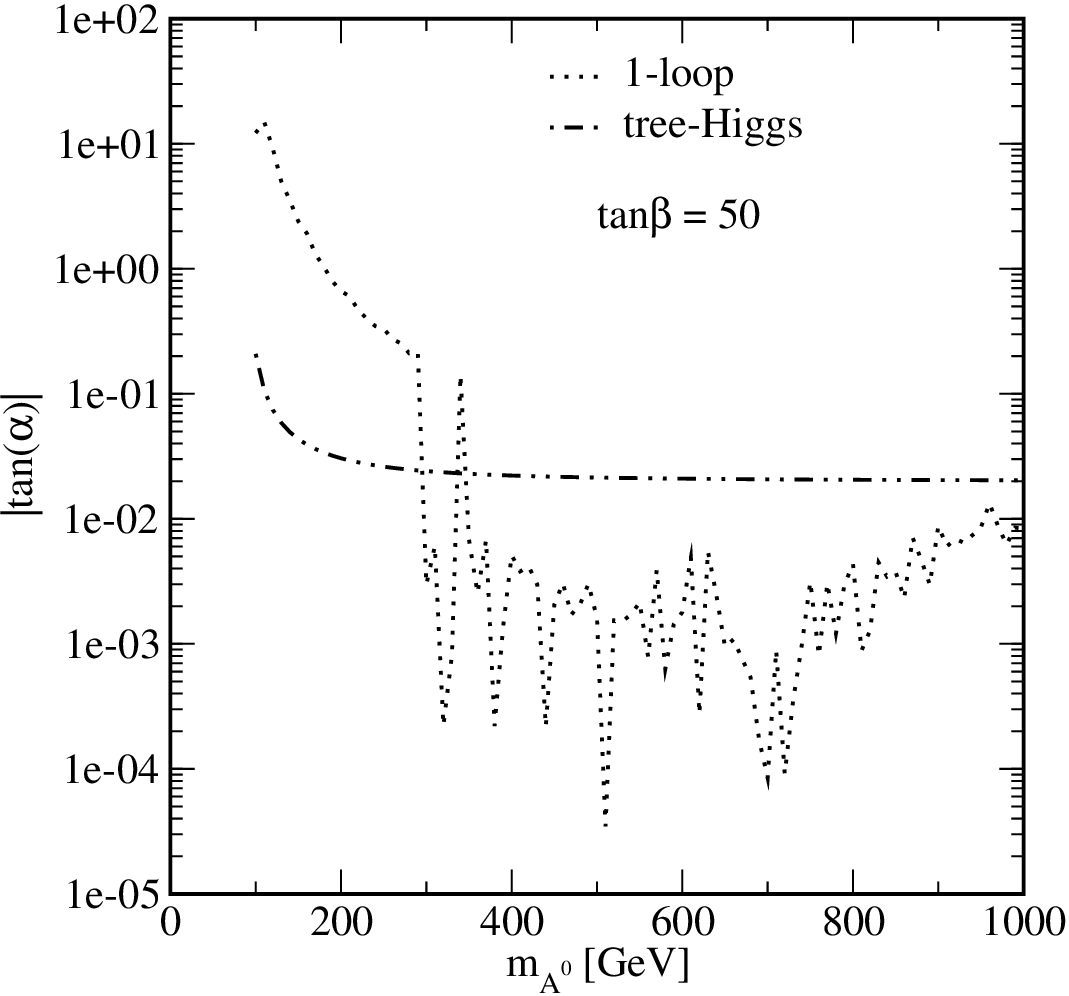}\\
(a) & (b)
\end{tabular}
\end{center}
    \caption{\textbf{a)} SUSY-EW contributions to the FCNC decay branching ratio
      $B(\hbs)$ as a function of $\mA$ for the MSSM parameters that maximize the various $\sigmapphbs$
      (cf. Fig.~\ref{fig:hbs_prod_ma}a);
    \textbf{b)} Absolute value of $\tan\alpha$ for the
    parameters that maximize the specific channel $\sigma(\pphzbs)$ where the ``small
    $\alpha_{\rm eff}$ scenario'' can be realized.}
    \label{fig:hbs_BR_ma}
\end{figure}
For a better understanding of these results, in
Fig.~\ref{fig:hbs_BR_ma}a we show the corresponding FCNC decay
branching ratio for the MSSM parameters that maximize the various
$\sigmapphbs$. In Fig.~\ref{fig:hbs_BR_ma}b, we focus on an
important feature of the particular $\PHiggslightzero$ channel, the
so-called ``small $\alpha_{\rm eff}$
scenario''~\cite{Carena:1999bh}. We plot there the (one-loop
corrected) value of $|\tan\alpha|$ for the parameters that maximize
the cross-section $\sigma(\pphzbs)$\,\footnote{The values of
$|\tan\alpha|$ shown in
  Fig.~\ref{fig:hbs_BR_ma}b correspond to the maximization of the
  SUSY-EW contributions in Fig.~\ref{fig:hbs_prod_ma}a. The
  corresponding values
  obtained from the maximization of the SUSY-QCD contributions
  (Fig.~\ref{fig:hbs_prod_ma}b) are essentially the same, except for the random
  fluctuations appearing in a Monte-Carlo computation.}.
The significant decrease of $|\tan\alpha|$ for $\mA\gtrsim 300\,GeV$
explains the relative stabilization of this cross-section in that
region (cf. Fig.~\ref{fig:hbs_prod_ma}a). We have also included the
corresponding curve in which the angle $\alpha$ is computed at the
tree-level (labelled \textit{tree-Higgs}).

In Table~\ref{tab:hbs-maxims} we summarize the numerical values of
$\sigmapphbs$, together with the parameters that maximize the
production rate for $\tb=50$ at the particular point
$\mA=200\,\GeV$. We also provide the value of $B(\hbs)$ and
$\Gamma(h\to X)$ at the maximum of the FCNC cross-section, and the
corresponding value of $B(\bsg)$.
\begin{table}
    \center
    \begin{tabular}{|c||c|c|c|}
        \hline
        $\,$ &  $H^0$ & $h^0$ & $A^0$ \\\hline\hline
        \sigmapphbs & $0.075\;\picobarn$ & $0.20\;\picobarn$ & $0.062\;\picobarn$ \\\hline
        events/$100\invfb$ & $7.5\times 10^3$ & $2.0\times 10^4$ & $6.2\times 10^3$ \\\hline
        $B(h\to bs)$ & $1.26\times 10^{-4}$ & $8.7\times 10^{-5}$ & $1.5\times 10^{-4}$ \\\hline
        $\Gamma(h\to X)$ & $9.5\,\GeV$ & $1.7\,\GeV$ & $11.3\,\GeV$ \\\hline
        $\tan\alpha$ & $-0.30$ & $0.67$ & $-0.01$ \\\hline
        $\msq$ & $880\,\GeV$ & $660\,\GeV$ & $990\,\GeV$ \\\hline
        $\mu$ & $1000\,\GeV$ & $970\,\GeV$ & $1000\,\GeV$ \\\hline
        $A_b$ & $-2350\,\GeV$ & $1900\,\GeV$ & $100\,\GeV$ \\\hline
        $A_t$ & $-1550\,\GeV$ & $-950\,\GeV$ & $2000\,\GeV$ \\\hline
        \Bbsg & $2.1\times 10^{-4}$ & $2.1\times 10^{-4}$ & $2.1\times 10^{-4}$ \\\hline
    \end{tabular}
    \caption{Maximum MSSM value of $\sigmapphbs$ (and of the number of
      $\bs$ events per $100\,\invfb$) at the LHC,  for $\mA=200\,\GeV$ and
      $\tan\beta=50$ under the assumption that the SUSY-QCD effects are negligible.
      Shown are also the corresponding values of the
      relevant branching ratio $B(\hbs)$ and of the total width of
      the Higgs bosons ($\PH\equiv \higgses$),
      together with the values of the SUSY
      parameters. The last row includes $B(\bsg)$, which is seen to lie in the allowed experimental range.}
    \label{tab:hbs-maxims}
\end{table}

Clearly, while the $\PHiggsheavyzero$ and $\PHiggspszero$ channels
follow a simple monotonous behavior, the particular
$\PHiggslightzero$ one behaves in a complex way. Let us further
elaborate on this point. At low values of $\mA\lesssim 300\,\GeV$,
$|\tan\alpha|$ is large and in this region the radiative corrections
can make it even larger (Fig.~\ref{fig:hbs_BR_ma}b). In this
situation, the coupling $\hz\Pb\bar{\Pb}\sim\sin\alpha/\cos\beta$
can be enhanced and the dominant production subprocess is
$\sigma(pp\to \hz \Pb\bar{\Pb})$. Note that, in this scenario,
$\Gamma(\hz\to \Pb\bar{\Pb})$ is also enhanced whereas
$B(\hz\to\bs)$ is suppressed (cf. Fig.~\ref{fig:hbs_BR_ma}a). The
net outcome is that the increase of the cross-section overcomes the
suppression of the branching ratio and the final result is one order
of magnitude larger than the \textit{tree-Higgs} expectation -- see
Fig.~\ref{fig:hbs_prod_ma}a. At large values of
$\mA\gtrsim300\,\GeV$, instead, where the renormalized value of
$\alpha$ (i.e. the effective  $\alpha_{eff}$) is significantly
smaller than the tree-level prediction\,\cite{Carena:1999bh}(cf.
Fig.~\ref{fig:hbs_BR_ma}b), the efficiency of the production
subprocess $pp\to \hz \Pb\bar{\Pb}$ is severely hampered. The Higgs
boson production is then dominated by gluon fusion at one loop
($gg\to \hz$) and since this mechanism is controlled by the
top-quark coupling $\hz\Pt\bar{\Pt}\sim\cos\alpha/\sin\beta$, it
becomes insensitive to variations in small $\alpha_{eff}$. Adding
this to the fact that $\Gamma(\hz\to \Pb\bar{\Pb})$ is strongly
suppressed and $B(\hz\to\bs)$ is correspondingly enhanced (cf.
Fig.~\ref{fig:hbs_BR_ma}a), we end up with a regime in which the
cross-section for $\PHiggslightzero\to\bs$ production is the most
favored one over a fairly sustained range of $\mA$.

The same effect was observed for the SUSY-QCD
contributions~\cite{Bejar:2005kv}, where the production rate is
augmented some three orders of magnitude with respect to the
\textit{tree-Higgs} case (cf. Fig.~\ref{fig:hbs_prod_ma}b). The
reason for this is that the SUSY-QCD effect on $\Gamma(\Phz\to \bs)$
is proportional to $\cos^2(\alpha-\beta)$ (see e.g. Eq.~(3.5) from
Ref.~\cite{Bejar:2004rz}). In the \textit{tree-Higgs} case, this
factor goes rapidly to zero for large $\mA$ because
$\alpha\to\beta-\pi/2$. However, at one loop such factor becomes
$\cos^2(\alpha_{eff}-\beta)$ and the previous relation does no
longer hold, so there is no such suppression. Another interesting
feature to note is that the SUSY-EW contribution to $\Gamma(\Phz\to
\bs)$ is roughly one order of magnitude smaller than the SUSY-QCD
one. This property is reflected at the level of $\sigma(\pphzbs)$
(Figs.~\ref{fig:hbs_prod_ma}a,b) because in this regime the leading
Higgs boson production mechanism is the same (gluon fusion) in both
cases. Using our analytical approximations, we can provide a simple
explanation for this numerical difference. If, for the sake of this
estimate, we take all SUSY masses of the same order, the ratio
between the SUSY-QCD and the SUSY-EW contributions to
$\Gamma(\Phz\to \bs)$ is expected to be
\begin{equation}\label{ratio}
\sim
10^2\,\left(\frac{\alpha_s}{\alpha_W}\,\frac{\delta_{23}}{|V_{ts}|}\right)^2\,
\left(\frac{m_W}{m_t}\right)^4\,\frac{1}{\tan^2\beta}\,\frac{\cos^2(\alpha_{eff}-\beta)}{\sin^2\alpha_{eff}}\,.
\end{equation}
Here we have used the effective SUSY-QCD-induced
$h^0\,b\,s$-coupling from Eq.~(3.5) of Ref.~\cite{Bejar:2004rz} and
the SUSY-EW contribution from the WF chargino loop effects, the
latter being similar to Eq.\,(\ref{eq:gamma_cha_bs1}) with
$\cos\alpha_{eff}$ replaced by $\sin\alpha_{eff}$. The prefactor
$\sim 10^2$ comes from the numerical factors $(2/3)^2\,16^2$
appearing in these formulae. For $\tan\beta=50$ and the numerically
determined values of $\delta_{23}\sim 10^{-1.5}\simeq 0.03$ and
$\alpha\sim 10^{-3}$ (corresponding to large $\mA>300\,\GeV$), we
find that (\ref{ratio}) is indeed of order $10$, as confirmed by
comparison of plots (a) and (b) in Fig.~\ref{fig:hbs_prod_ma}. This
explains nicely the approximate numerical relation between the
SUSY-QCD and SUSY-EW effects in the ``small $\alpha_{\rm eff}$
scenario'' from simple dynamical considerations. {Finally, let us
mention that, for very large values of $\mA$ (namely $\mA \gtrsim
800\,\GeV$) the small $\alpha_{eff}$ scenario is no longer
maintained, and so the tree-level and 1-loop values for $\alpha$
tend to merge (see Fig.~\ref{fig:hbs_BR_ma}). By the same token,
also the two $\sigma(\HepProcess{\Pproton\Pproton \HepTo
\PHiggslightzero \HepTo bs})$ curves, those for $\alpha$ being
computed at the tree-level and at 1-loop respectively (cf.
Fig.~\ref{fig:hbs_prod_ma}), tend to approach each other in the
aforementioned limit\footnote{A similar discussion, specifically for
the SUSY-QCD case, can be found in Ref.~\cite{Bejar:2005kv}.}. It is
worth recalling that the light Higgs mass ($m_{\PHiggslightzero}$)
reaches its upper bound ($m_{\PHiggslightzero}\lesssim 130\,GeV$)
for $\mA \to \infty$. Since $\alpha$ also reaches a constant value
for very large $\mA$, we can understand the reason why
$\sigma(\HepProcess{\Pproton\Pproton \HepTo \PHiggslightzero \HepTo
bs})$ exhibits an almost flat slope in this asymptotic regime.}

Let us now evaluate the $\bs$ event rate attained from the direct
production mechanism $\sigma(\pphbsD)$. In the MSSM case, we may
have both SUSY-EW and SUSY-QCD effects. The number of diagrams is
rather large and, as mentioned above, it can be inferred from those
in Ref.\,\cite{LopezVal:2007rc} after appropriate replacements of
the internal and external lines. Let us first concentrate upon the
purely SUSY-EW part. In Table~\ref{tab:directbs} we present, in a
nutshell, the predicted values for the direct production of $bs$
pairs through gluon-gluon fusion $\sigma(\ppggbs)$ from the
electroweak charged-current effects (chargino and charged Higgs
boson loops). The computation of $\sigma(\ppggbs)$ is performed
within the parameter set that optimizes the $\bs$ production rate
through $\pphzbs$ (central column of Table~\ref{tab:hbs-maxims}). We
notice that the non-standard contributions are tiny (of order
$10^{-5}$ at most) and, moreover, a destructive interference arises
when we add up the chargino and charged-Higgs boson mediated
amplitudes. The latter are suppressed by small factors of
$m_W^2/m^2_H, \,m_W^2/M^2_{SUSY}$ as compared to the SM ones. In
addition, the GIM mechanism involved in the SM part is much less
severe in the down-like quark sector as compared to its effect in
the up-quark sector owing to the presence of the large top quark
mass in the latter case.

\begin{table}
\centerline{\begin{tabular}{|c||c|} \hline partial contribution &
$\sigma(\ppggbs) (\picobarn)$ \\ \hline \hline Charged Higgs&
$9.7\times 10^{-6}$ \\ \hline Chargino & $1.1\times 10^{-5}$ \\
\hline SUSY-EW & $5.7\times 10^{-7}$\\ \hline MSSM & $1.3\times
10^{-3}$ \\ \hline SM & $1.3\times 10^{-3}$\\ \hline
\end{tabular}}
\caption{Different SUSY-EW contributions to the direct $\bs$
production for the choice of parameters that maximize
$\sigma(\pphzbs)$ (central column of Table~\ref{tab:hbs-maxims}). In
the absence of significant SUSY-QCD effects, the SM and the overall
MSSM contributions are coincident.} \label{tab:directbs}
\end{table}

\separ Unfortunately, the probability to observe ``direct $\bs$
events'' does not improve in the presence of explicit sources of
supersymmetric flavor mixing. In fact, our calculation of the
SUSY-QCD effects on the direct $\bs$ production shows that the
corresponding cross-section cannot compete with the FCNC Higgs boson
decay channels. The reason is twofold: i) the mass insertions
arising in the gluino and neutralino loops produce vertices of the
guise $(\tilde{g},\chi^0_{\alpha})\,\Pbottom\APstrange \sim m_b(A_b
- \mu\,\tan\beta)$ (cf. Section \ref{sec:higgs-mediated}), and so
proportional to $m_b$, whereas in the $\tc$ channel the effective
vertices get a factor of $m_t$; ii) the particular combination $A_b
- \mu\,\tan\beta$ becomes directly and stringently constrained by
the experimental data on $B(\bsg)$, which impose small values of
$\delta_{23}$ or very heavy gluino/neutralino masses. For example,
if we take the set of MSSM parameters for which the
$\PHiggsheavyzero$-mediated SUSY-QCD case is optimized (see Table 1
of Ref.\,\cite{Bejar:2005kv}) we find $\sigma(\pphbsD)=1.1 \times
10^{-4}\, \picobarn$, versus $0.45\, \picobarn$ from the
corresponding Higgs decay mechanism. These results can be compared
with the purely SM contribution to direct $\bs$ production, which we
find it to be $\sigma(\pphbsD)_{\rm SM} = 1.3 \times 10^{-3}\,
\picobarn$ and  entails $\sim 100$ events per $\int {\cal
L}\,dt=100\,\invfb$. We conclude that the enhancement capabilities
of the $\bs$ events within the MSSM are dominated by the FCNC Higgs
boson decay modes rather than by the direct FCNC production
processes.

\section{Numerical analysis of the $\tc$ production rate}
\label{sec:analysis-tc}

\begin{figure}[tp]
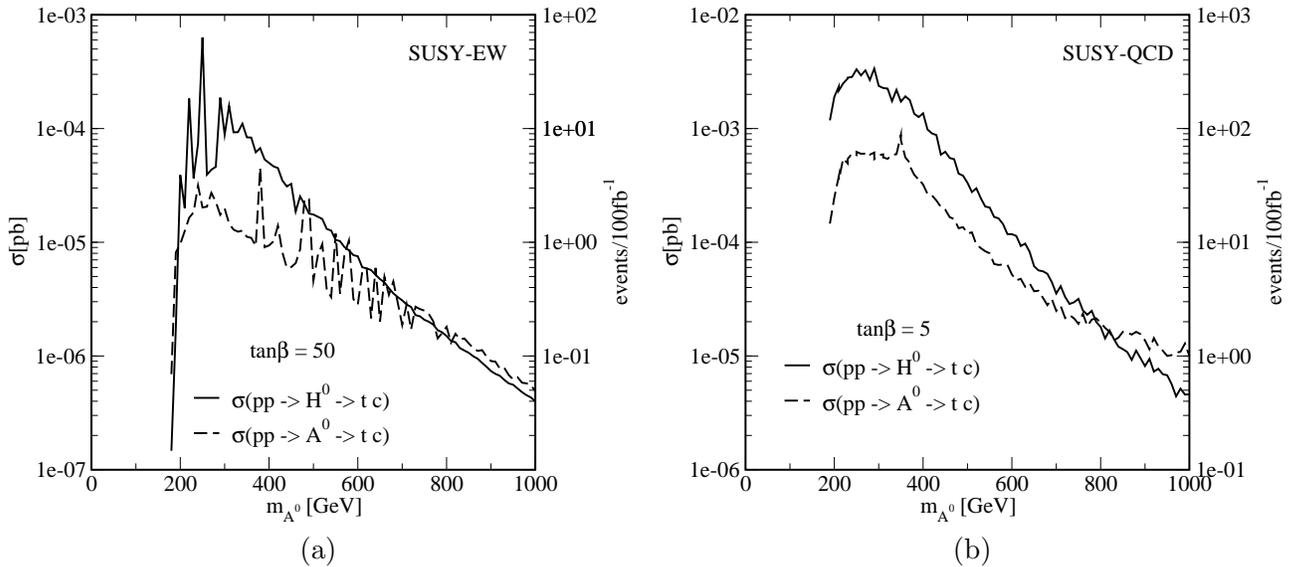

    \begin{center}
    \begin{tabular}{cc}
        \includegraphics*[height=0.42\textwidth]{htc_prod_pphqq_ma} &
        \includegraphics*[height=0.42\textwidth]{htcSQCD_prod_pphqq_ma} \\
        (a) & (b)
    \end{tabular}
    \end{center}
    \caption{Maximum contributions to $\sigmapphtc$ -- cf.
      Eq.\,(\ref{eq:hqq-def})-- as a function of
      $\mA$ (at fixed $\tb$) taking into account \textbf{a}) SUSY-EW interactions
      with minimal flavor-mixing and \textbf{b}) SUSY-QCD interactions without electroweak effects (see
      Ref. \cite{Bejar:2005kv}).}
    \label{fig:htc_prod_ma}
\end{figure}
Let us now present the results of an equivalent maximization
procedure for the $\tc$ production rate, $\sigmapphtc$. In
Fig.~\ref{fig:htc_prod_ma}a, we plot the maximum cross-sections
attained at different values of the CP-odd Higgs boson mass, $\mA$.
For better comparison, in Fig.~\ref{fig:htc_prod_ma}b we show the
corresponding maximization for the SUSY-QCD
case\,\cite{Bejar:2005kv}. As in the previous $\bs$ analysis, the
$\tc$ production rate becomes again favored in the large $\tan\beta$
regime: this is due to the correlations between the Higgs boson
production and its subsequent decay (see the analytical estimates
for the $\tc$ channel in Section \ref{sec:higgs-mediated}). Further
numerical details of the optimal MSSM configuration are quoted in
Table.~\ref{tab:htc-maxims}. We find that the predicted rates are
rather inconspicuous (of the order of $\sigma \sim
10^{-4}\,\picobarn$ at most), and hence difficult to detect.  We
emphasize that this result is essentially triggered by the
charginos. Neutralinos, again, have an even lesser impact. Indeed,
if we recall the analytical estimates in
Section~\ref{sec:higgs-mediated} and the results from
Tables~\ref{tab:hbs-maxims} and \ref{tab:htc-maxims}, we observe
that the optimal value of $A_t$  in the latter (which is responsible
for the neutralino contribution to the $\tc$ channel) is much
smaller than the optimal value of $A_b$ in the former.

It is instructive to trace the main differences between the SUSY-EW
effects on the two channels ($\tc$ and $\bs$) by using simple
qualitative arguments based on the dynamical features of the MSSM.
Fortunately, this can be immediately done from the analytical
estimates presented in Section~\ref{sec:higgs-mediated} and the
optimal parameter sets in each case, which can be extracted from
Tables~\ref{tab:hbs-maxims} and \ref{tab:htc-maxims}. We note that
the values of $\tan\alpha$ and $\tan\beta$ are quite similar in both
cases. Therefore, the production cross-section obtained from the
FCNC decays of the MSSM Higgs bosons must render essentially the
same result and cancels in the ratio.
\begin{table}
    \center
    \begin{tabular}{|c||c|c|}
        \hline
        $h$ &  $H^0$ & $A^0$ \\\hline\hline
        \sigmapphtc & $8.8\times 10^{-5}\;\picobarn$ & $2.0 \times 10^{-5}\;\picobarn$ \\\hline
        events/$100\invfb$ & $8.8$ & $2.0$ \\\hline
        $B(h\to tc)$ & $8.5\times 10^{-7}$ & $2.4\times 10^{-7}$ \\\hline
        $\Gamma(h\to X)$ & $36\,\GeV$ & $39\,\GeV$ \\\hline
        $\tan\alpha$ & $0.046$ & $-0.11$ \\\hline
        $\msq$ & $300\,\GeV$ & $350\,\GeV$ \\\hline
        $\mu$ & $350\,\GeV$ & $350\,\GeV$ \\\hline
        $A_b$ & $-675\,\GeV$ & $-1000\,\GeV$ \\\hline
        $A_t$ & $20\,\GeV$ & $-75\,\GeV$ \\\hline
        \Bbsg & $2.9\times 10^{-3}$ & $2.5\times 10^{-3}$ \\\hline
    \end{tabular}
    \caption{Maximum SUSY-EW induced value of $\sigmapphtc$ (and of the number of
      $\tc$ events per $100\,\invfb$) at the LHC,  for $\mA=300\,\GeV$ and
      $\tan\beta=50$. Shown are also the corresponding values of the
      relevant branching ratio $B(\hbs)$ and of the total width of
      the Higgs bosons, together with the values of the SUSY
      parameters. The last row includes $B(\bsg)$.}
    \label{tab:htc-maxims}
\end{table}
Taking advantage of this fact, the ratio of the two cross-sections
for $h=H^0,A^0$ (the two states available for both $\bs$ and $\tc$
final states) is essentially given by that of the branching ratios
and also by that of the corresponding partial widths,
\begin{eqnarray}
\frac{\sigmapphbs}{\sigmapphtc} \sim \frac{B(\hbs)}{B(\htc)}=
\frac{\Gamma(\hbs)}{\Gamma(\htc)} \label{eq:aprox}\,.
\end{eqnarray}
In the specific case of the heavy CP-even neutral Higgs boson, the
approximate analytical form is obtained from
Eqs.(\ref{eq:gamma_cha_bs1}-\ref{eq:gamma_cha_tc2}). Plugging the
MSSM parameters corresponding to the optimal configuration for each
of the channels, we realize that the dominant contribution to $\bs$
comes from the chargino-mediated WF corrections, while the vertex
corrections drive the main part of the $\tc$ one. Therefore, we can
approximate the ratio of the overall cross-sections of
Eq.~(\ref{eq:aprox}) as follows:
\begin{eqnarray}
\frac{\sigmapphbs}{\sigmapphtc} \sim
\left(\frac{m_t}{m_b}\right)^2\,\left(\frac{A_t}{A_b}\right)^2 \sim 10^3
\label{eq:compar}.
\end{eqnarray}
which is in good agreement with the 3 orders of magnitude that arise
from the exact numerical computation, {once we plug the
corresponding values of the quark masses and the trilinear
couplings, $A_t$ from Table~\ref{tab:hbs-maxims} and $A_b$ from
Table~\ref{tab:htc-maxims}}. Therefore, we confirm that, in the
optimal scenario, the difference between both channels can be well
accounted for by the quark mass insertions squared times the ratio
squared of the trilinear couplings  (top quark mass and $A_t$, in
the $\bs$ channel, versus bottom quark mass and $A_b$, in the $\tc$
channel).

In comparison, the situation for the direct SUSY flavor-changing
production of $\tc$ pairs is remarkably distinct. As we have argued
in Section~\ref{sec:analysis-bs}, the enhancement factor of the
relevant couplings is much larger in this case, being now
proportional to the top quark mass $m_t$. In addition, the $\bsg$
constraints can be more easily eluded (just by an appropriate choice
of the MSSM parameters, see Refs.~\cite{\singletoppro, \singletop}).
Numerically speaking, the direct SUSY-EW production of $\tc$ pairs
furnishes $\sigma \sim 0.01\,\picobarn$, or $\sim 10^3$ events per
$\int{\cal L}\,dt=100$ $\invfb$ -- see Ref.~\cite{\singletop} for a
comprehensive discussion including the direct SUSY-QCD effects. Such
optimal scenarios are favored by relatively low supersymmetric mass
scales, namely $M_{SUSY} \sim 250\, \GeV$, $M_1, M_2 \sim 100\,\GeV$
(for the electroweak soft gaugino masses), together with large
flavor-mixing values of $ \delta_{23}$. The compliance with the
$\bsg$ constraints is achieved by balancing the different SUSY-EW
pieces involved in this process, i.e. assuming relatively light
masses for charginos, neutralinos, charged Higgs bosons and the top
squark, and also moderate values of $\tan\beta$. The dominance of
the direct $\tc$ production mechanism holds also in the case of
SUSY-QCD contributions~\cite{\singletop}. In the most favorable
circumstances, one can reach $\sigma \sim 1\,\picobarn$ from direct
production, whereas the optimal Higgs boson-mediated output lies
around $\sigma \sim 10^{-3}\,\picobarn$ (cf.
Fig.~\ref{fig:htc_prod_ma}b).

\section{Discussion and conclusions}
\label{sect:conclusions}

A number of studies have been devoted to the analysis of the FCNC
signatures carried by electrically neutral pairs of heavy quarks of
different flavors as an strategy to unravel hints of New Physics in
the forthcoming LHC data. These events are very rare in the SM and,
therefore, their observation could be highly revealing. Some of
these studies, including ours, have focused on the possibility that
the underlying new physics could be Supersymmetry, in particular the
unconstrained MSSM. In this letter, we have discussed the SUSY
effects on the production and subsequent FCNC decay of the neutral
MSSM Higgs bosons ($\PH = \higgses$) into heavy quark pairs $qq'
=\bs,\tc$ at the LHC, i.e. $\pphqq$, and we have compared the
results with the direct FCNC production mechanism $\pphqqD$.
Furthermore, we have computed the SUSY-EW corrections to
$\sigma(\pphqq)$ and also the SUSY-QCD and SUSY-EW contributions to
$\sigma(\pphbsD)$. These results extend the analyses previously
presented for the SUSY-QCD effects on
$\sigma(\pphqq)$\,\cite{Bejar:2005kv} and the SUSY-QCD and SUSY-EW
ones on
$\sigma(\pphtcD)$\,\cite{Liu:2004bb,Guasch:2006hf,Eilam:2006rb,LopezVal:2007rc}.
At the end of the day, we have nicely completed the map of MSSM
predictions for the heavy $qq'$-pairs produced at the LHC.

\begin{table}
\begin{center}
\begin{tabular}{|c||c|c|} \hline
FCNC mechanism \ channel & $\Ptop -\Pcharm$ & $\Pbottom -\Pstrange$ \\
\hline \hline
\multicolumn{3}{|c|}{Higgs decay-mediated mechanism} \\
\hline \hline
SUSY-QCD & $\sim 10^{-3}\,\picobarn$ & $\sim 1 \,\picobarn $ \\
\hline
SUSY-EW  & $\sim 10^{-4}\,\picobarn$ & $\sim 10^{-1}\,\picobarn$ \\
\hline\hline
\multicolumn{3}{|c|}{Direct FCNC production mechanism}  \\
\hline \hline
SUSY-QCD dominance & $\sim 1 \,\picobarn$ & $\sim 10^{-3}\,\picobarn$ \\
\hline
SUSY-EW dominance & $\sim 10^{-2}\,\picobarn$ & $\sim 10^{-4}\,\picobarn$ \\
\hline
SM & $\sim 10^{-7}\,\picobarn$ & $\sim 10^{-3}\,\picobarn$ \\
\hline
\end{tabular}
\caption{Summary of optimal SUSY contributions for $\tc$ and $\bs$
events at the LHC, within order of magnitude. For the direct
production, we follow the same procedure as in
\cite{LopezVal:2007rc}, viz. we explore scenarios where there is a
dominant SUSY component (SUSY-EW or SUSY-QCD) and allow a smaller
contribution from the other. In the last row, we quote the SM
prediction.} \label{tab:vs}
\end{center}
\end{table}
The upshot of this lengthy investigation, which includes also the
analyses from the previous works
\cite{Bejar:2004rz,Bejar:2005kv,Guasch:2006hf,LopezVal:2007rc}), is
summarized in  Table~\ref{tab:vs}. These numerical results are
obtained in full consistency with the stringent experimental
constraints from $\bsg$. The most favorable channels turn out to be
the following: 1) the Higgs boson FCNC decays into $\bs$, and 2) the
direct production of $\tc$ pairs, both of them with maximal
cross-sections of $\sim 1\picobarn$ and dominated by SUSY-QCD
effects. In that table, we also collect the results
$\sigma(\pphbsD)_{\rm max}\sim 10^{-3}\picobarn$ from SUSY-QCD and
$\sim 10^{-4}\picobarn$ from SUSY-EW. Clearly, the direct production
of $bs$ pairs in the MSSM is highly inefficient as compared to the
rate that could originate from FCNC decays of the MSSM Higgs bosons.
Finally, if the SUSY-QCD part is negligible (e.g. because the
gluinos are very heavy or the flavor-mixing terms are too small),
the SUSY-EW loops alone yield a small $\tc$ rate from Higgs boson
decays, $\sigma(\pphtc)_{\rm max}\sim 10^{-4}\picobarn$.
Nonetheless, in this case, we also have $\sigma(\pphbs)_{\rm
max}\sim 0.1\picobarn$, implying $\sim 10^4$ $\bs$ pairs per
$100\,\invfb$ of integrated luminosity. In other words, in the
absence of strong supersymmetric interactions, the $tc$ signature
disappears for all practical purposes but we could still count on a
fairly large amount of  $bs$ events triggered by electroweak
supersymmetric sources.

Despite the predicted number of FCNC $q\,q'$ events is sizeable in
some cases, it is far from obvious that they could be efficiently
disentangled from the underlying background of QCD jets where they
would be immersed, even in the most favorable conditions. For
example, it is well known that the simple two-body decay $\PH\HepTo
\Pbottom\,\APbottom$ is virtually impossible to isolate, due to the
huge irreducible QCD background from $\Pbottom\,\APbottom$ dijets.
This led, long time ago, to complement the search with many other
channels, particularly with the radiative decay
$\PH\HepTo\Pphoton\,\Pphoton$, which has been identified as an
excellent signature in the appropriate range\,\cite{\Hunter}.
Similarly, the FCNC Higgs boson decay channels may help to
complement the general Higgs boson search strategies, mainly because
the FCNC processes should be essentially free of QCD background.
Notwithstanding, other difficulties can appear related to the
misidentification of jets. For instance, for the $\bs$ final states
misidentification of $\Pbottom$~quarks as $\Pcharm$~quarks in
$\Pcharm\Pstrange$-production from charged currents may obscure the
possibility that the $\bs$ events can be really attributed to Higgs
boson FCNC decays. This also applies to the $\tc$ final
states~\cite{Stelzer:1998ni}, where misidentification of
$\Pbottom$~quarks as $\Pcharm$~quarks in e.g. $\Ptop\Pbottom$-
production, might be a source of background to the $\tc$ events,
although in this case the clear-cut top quark signature should be
much more helpful (specially after performing a study of the
distribution of the signal versus the background). These studies,
however, go beyond the scope of the analysis presented in this work.

To summarize, in this letter we have completed the calculations
necessary to account for the FCNC-triggered $qq'$ events that could
emerge at the LHC from supersymmetric sources in the unconstrained
MSSM. While admitting that the practical detection of these events
can be difficult, the message from the theoretical side seems now
crystal-clear: the MSSM has the potential to provide large amounts
of heavy quark pairs from genuine supersymmetric FCNC interactions,
to wit: in the $\bs$ channel, the production and subsequent FCNC
decay of a neutral Higgs boson entails potentially large
enhancements (mainly in the large $\tan\beta$ regime) which could
boost the predicted cross-sections up to $\sigma(\pphbs)\sim
1\,\picobarn$ from SUSY-QCD, and $\sim 0.1\,\picobarn$ from SUSY-EW,
whilst the direct production is not enhanced at all with respect to
the SM result. At variance with this situation, the $\tc$ channel
can be maximally enhanced from the direct mechanism, whereas the
Higgs boson-mediated rate $\sigma(\pphtc)$ proves to be negligible.
From the SUSY-QCD side, we find the direct production cross-section
$\sigma(\pptc)_{\rm max} \sim 1\,\picobarn$, while from SUSY-EW we
obtain $\sigma_{\rm max}(\pptc) \sim 0.01\,\picobarn$, the
respective number of events being rather large: $10^5$ and $10^3$
$\tc$ pairs per $\int{\cal L}\,dt=100\,\invfb$. The $\tc$ signature
should obviously be the preferred one for a more promising
experimental tagging owing to the presence of the top quark in the
final state.

\section*{Acknowledgements}
The work of SB has been supported by European Community's Marie-Curie Research
Training Network under contract MRTN-CT-2006-035505 `Tools and Precision
Calculations for Physics Discoveries at Colliders' ; DLV by the MEC FPU grant
Ref. AP2006-00357; JG and JS in part by MEC and FEDER under project
FPA2007-66665
and also by DURSI Generalitat de Catalunya under
project 2005SGR00564. This work has also been supported by the Spanish
Consolider-Ingenio 2010 program CPAN CSD2007-00042.

\providecommand{\href}[2]{#2}

\end{document}